# COULOMB CORRECTIONS TO SUPERALLOWED BETA DECAY IN NUCLEI.

N. Auerbach

School of Physics and Astronomy, Tel Aviv University, Tel Aviv 69978, Israel
and TRIUMF, 4004 Wesbrook Mall, Vancouver, B.C., Canada, V6T 2A3

Abstract : Corrections to the superallowed beta decay matrix elements are evaluated in perturbation theory using the notion of the isovector monopole resonance. The calculation avoids the separation into different contributions and thus presents a consistent, systematic and more transparent approach. Explicit expressions for $\delta_c$ as a function of the mass number $A$ are given.

## I INTRODUCTION

One of the recent activities in nuclear structure are the attempts to determine the corrections one has to introduce in the evaluation of the beta-decay matrix elements for super-allowed transitions in $T=1$, $T_z=+1$ (or $T_z=-1$) nuclei [1,2].
This is considered to be an important issue because using the measured ft values one can relate these to the u-quark to d-quark transition matrix element (m.el.) $V_{ud}$ in the Cabibbo-Kobayashi-Maskawa (CKM) matrix. In the Standard Model (SM) this matrix satisfies the unitarity condition, that is the sum of squares of the matrix elements in each row (column) is equal to one :

$$V_{ud}^2 + V_{us}^2 + V_{ub}^2 = 1 \qquad (1)$$

Departures from 1 may indicate physics not described by the SM.
In order to use the experimental ft values to determine $V_{ud}$ one has to introduce corrections [1-3]. There is a class of important radiative corrections which we will not treat here. Discussions of these can be found abundantly in the literature [1-3].
The second type of correction, that is usually termed as the isospin symmetry breaking term, denoted as $\delta_c$ and defined by the following equation:

$$|M_F|^2 = |M_F^0|^2 (1-\delta_c) \qquad (2)$$

where $M_F$ is the physical Fermi matrix element:

$$M_F = \langle \Psi_1 | T_+ | \Psi_2 \rangle \qquad (3)$$

$|\Psi_1\rangle$ and $|\Psi_2\rangle$ are the parent and daughter physical states. The symbol $M_F^0$ stands for the Fermi matrix element obtained in the limit when in the Hamiltonian all the charge-dependent parts are put to zero, and the wave functions are eigenstates of the charge-independent Hamiltonian.

The calculation of $\delta_c$ is usually done by breaking it up into several contributions that result from the inclusion of charge-dependent terms in the Hamiltonian of the nucleus. This separation into different types of contributions is model dependent. As pointed out recently [4] the approach taken in a number of studies [1, 2] used the notion of analog spin [5], (also called the $W$-spin), instead of isospin, and this complicates matters because in eq. (2) the isospin raising operator appears and not the $W_+$.

In the present approach we start from a charge-independent Hamiltonian so that the matrix element in eq. (1) is exactly $\sqrt{2T}$ and we then treat the Coulomb force in perturbation theory. In the way we approach the problem there is no need to break up the contribution of the Coulomb interaction into various separate components. All the effects of Coulomb mixing (such as isospin mixing, the change in the radial part of the wave functions, etc) are taken into account in a single term. (Some aspects of this approach have already been presented in the past [6], [7]).

## II COULOMB MIXING

We start by introducing a nuclear charge independent Hamiltonian which does not contain any charge dependent parts- $H_0$. The eigenstates of this Hamiltonian with isospin $T$ and $T_z$ will be denoted as $|T,T_z\rangle$ and:

$$H_0|T,T_z\rangle = E_T|T,T_z\rangle \qquad (4)$$

The $2T+1$, components with different $T_z$ values are degenerate.

The action of the isospin lowering and raising operators, $T_-$, $T_+$, gives:

$$T_-|T,T\rangle = \sqrt{2T}|T,T-1\rangle \; ; \; T_+|T,T-1\rangle = \sqrt{2T}|T,T\rangle \qquad (5)$$

We now add to the charge independent Hamiltonian a charge dependent part $V_{CD}$.

The dominant part in the charge dependent interaction is the charge asymmetric Coulomb force $V_C$. (While the charge-dependent components of the two-body nuclear force might be important in changing the relative spacing of levels in the analog nucleus its influence on isospin mixing is expected to be small). In what follows we will deal only with off-diagonal matrix elements of the Coulomb interaction. Because of the long range nature of the Coulomb force, the prevailing part will be in such cases the one-body part.

As a good approximation we take the potential of a uniformly charged sphere.

Inside the sphere $(r \leq R)$:

$$V_C(r) = -\frac{Ze^2}{R^3} \sum_i \left(\frac{1}{2}r_i^2 - \frac{3}{2}R^2\right)\left(\frac{1}{2} - t_z(i)\right) \tag{6}$$

Of interest to us here is the isovector part of the potential. Any off-diagonal matrix element between two states of the isovector part is:

$$\langle 0|V_C|n\rangle = \frac{Ze^2}{2R^3}\langle 0|\sum_i r_i^2 t_z(i)|n\rangle \equiv \frac{Ze^2}{2R^3}\langle 0|M_0^{(1)}|n\rangle \tag{7}$$

where $M_0^{(1)}$ denotes the z-component of isovector monopole operator.
It obvious that if the state $|n\rangle$ is the giant isovector monopole state [8] then the above matrix element will exhaust much of the Coulomb sum rule [8].

We will now find in perturbation theory the effect of the charge-dependent part on the wave functions of the two members of the isomultiplet, $|T,T\rangle$ and $|T,T-1\rangle$.

$$\Psi_1 = \left(|T,T\rangle + \varepsilon_T|M_{T,T}\rangle + \varepsilon_{T+1}|M_{T+1,T}\rangle\right)N_1^{-1} \tag{8a}$$

$$\Psi_2 = \left(|T,T-1\rangle + \eta_{T-1}|M_{T-1,T-1}\rangle + \eta_T|M_{T,T-1}\rangle + \eta_{T+1}|M_{T+1,T-1}\rangle\right)N_2^{-1} \tag{8b}$$

where $|M_{T',T_z'}\rangle$, are the $T',T_z'$ components of the isovector monopole, and where

$$N_1 = \sqrt{1 + \varepsilon_T^2 + \varepsilon_{T+1}^2} \quad \text{and} \quad N_2 = \sqrt{1 + \eta_{T-1}^2 + \eta_T^2 + \eta_{T+1}^2} \tag{9}$$

The admixtures are given in perturbation theory by the equations:

$$\varepsilon_i = \frac{\langle T,T|V_C^{(1)}|M_{T+i,T}\rangle}{E_{M_{T+i,T}} - E_0}, \quad i = 0,1 \tag{10}$$

where $E_0$ is g.s. energy in this nucleus,

$$\eta_i = \frac{\langle T,T-1|V_C^{(1)}|M_{T+i,T-1}\rangle}{E_{M_{T+i,T-1}} - E_1}, \quad i = -1,0,1 \tag{11}$$

Here $E_1$ is the energy of the analog state.
One can write these as:

$$\varepsilon_i = \langle T,T,1,0|T+i,T\rangle \langle T+i\|V_C^{(1)}\|T\rangle /(E_{M_{T+i,T}} - E_0) \tag{12}$$

$$\eta_i = \langle T,T,1,0|T+i,T-1\rangle \langle T+i\|V_C^{(1)}\|T\rangle /(E_{M_{T+i,T-1}} - E_1) \tag{13}$$

The first bracketed expression is the Clebsch-Gordan (CG) coefficient, while the second bracket is the reduced matrix element. The reduced matrix element for large excess neutron (proton) nuclei is such that the components with lower isospin have larger values [8]. However we will deal with two- nucleon excess nuclei (T=1, states) and in this case the differences are very small, so we will assume that the various reduced matrix elements are equal. The energy denominators are the excitations of the isovector monopole components in the parent and daughter nuclei either with respect to the ground state or the analog state. We will denote these energies as, $\Delta E_M^i$.

The $T+i$ components are split by the symmetry potential:

$$V_s = \frac{V_1}{A}(\vec{t}\cdot\vec{T}) \tag{14}$$

Where $\vec{t}$ is the isospin operator of the isovector excitation.

For the various components of the monopole excitation:

$$E_{T+i}^s = \frac{V_1}{A}[(T+i)(T+i+1) - T(T+1) - 2] \tag{15}$$

and

$$\Delta E_M^i = \xi\,\hbar\omega + E_{T+i}^s \tag{16}$$

where $\hbar\omega = 41A^{-1/3}$ MeV and $\xi$ is a numerical factor which depends on the model used to describe the isovector monopole. The range of values for this parameter is between 3 and 4 [8].

Introducing the values of the CG coefficients and the notation $u$ for the reduced matrix element, denoting $\kappa = \dfrac{2V_1}{\xi\hbar\omega A}$ we can write:

$$\varepsilon_0 = -\sqrt{\frac{T}{T+1}}\frac{u}{\xi\hbar\omega}\frac{1}{(1-\kappa)} \tag{17}$$

$$\varepsilon_1 = \sqrt{\frac{1}{T+1}}\frac{u}{\xi\hbar\omega}\frac{1}{(1+T\kappa)} \tag{18}$$

$$\eta_{-1} = -\sqrt{\frac{2T-1}{T(2T+1)}}\frac{u}{\xi\hbar\omega}\frac{1}{[1-(T+1)\kappa]} \tag{19}$$

$$\eta_0 = -\frac{T-1}{\sqrt{T(T+1)}} \frac{u}{\xi\hbar\omega} \frac{1}{(1-\kappa)} \tag{20}$$

$$\eta_1 = \sqrt{\frac{4T}{(2T+1)(T+1)}} \frac{u}{\xi\hbar\omega} \frac{1}{(1+T\kappa)} \tag{21}$$

The matrix element of interest here is:

$$\langle\Psi_1|T_+|\Psi_2\rangle = \sqrt{2T}\left(1 + \varepsilon_0\eta_0 + \varepsilon_1\eta_1\sqrt{\frac{2T+1}{T}}\right) N_1^{-1} N_2^{-1} \tag{22}$$

where:

$$N_1 = \left(1 + \varepsilon_0^2 + \varepsilon_1^2\right)^{1/2} \tag{23a}$$

$$N_2 = \left(1 + \eta_{-1}^2 + \eta_0^2 + \eta_1^2\right)^{1/2} \tag{23b}$$

It is worthwhile at this point to note that in case of complete degeneracy of the $T+i$ components of the isovector monopole state (that is when $V_1 = 0$) the result for the physical matrix element in eq. (1) will be $\sqrt{2T}$, the same as in the case when there is no charge dependent perturbation. This occurs actually when isospin mixing is not zero, and not necessarily negligible. It happens because the parent and daughter states can be related to each other via the action of the $T_+$ operator in spite of the fact that the Hamiltonian is not charge-independent.

We will apply this formalism to the $T_z = \pm 1$ nuclei for which experimental results exist for number of nuclei. The value of $V_1 \approx 100$ MeV. The value of $\kappa$ is smaller than 1 even in nuclei as light as $^{10}C$ and it becomes very small for A=50. We can express all the above coefficients in terms of one, and we choose $\varepsilon_1$. Neglecting terms quadratic (or higher powers) in $\kappa$, we calculate the matrix element in eq.s (17-23) and arrive, after some algebra, at the expression:

$$\langle\Psi_1|T_+|\Psi_2\rangle^2 = 2T\left(1 - 2(T+1)\frac{V_1}{\xi\hbar\omega A}\varepsilon_1^2\right)^2 \tag{24}$$

Therefore to order $\varepsilon_1^2$, using the definition in eq. (2) we find:

$$\delta_c = 4(T+1)\frac{V_1}{\xi\hbar\omega A}\varepsilon_1^2 \tag{25}$$

and using the expression for $\hbar\omega$ in MeV,

$$\delta_c = 4(T+1)\frac{V_1}{41\xi A^{2/3}}\varepsilon_1^2 \tag{26}$$

$\varepsilon_1^2$ is the isospin admixture of the $T+1$ state into the $T, T_z = T$ ground state. This is commonly defined as isospin impurity. Note that $\delta_c < \varepsilon_1^2$ because the coefficient in front of $\varepsilon_1^2$ for the nuclei considered is less than 1. For large $T$ there is the factor $T+1$ that enhances the value but it is cancelled by the factor $(T+1)^{-1}$ in $\varepsilon_1^2$ [8].
For $T=1$ nuclei

$$\delta_c = 8\frac{V_1}{41\xi A^{2/3}}\varepsilon_1^2 \tag{27}$$

We will now use different models for $\varepsilon_1^2$ introduced in the past and presented in ref. [8]. We will use the value $\xi \approx 3$ and take $V_1 = 100$ MeV. Of course as we mentioned some models predict $\xi > 3$ and the value of $V_1$ is not well determined and values lower than 100 MeV are also used sometimes. However both uncertainties will not change our main conclusions.

### III  RESULTS

The isospin impurities in the ground state of a nucleus (or in its isobaric analog) are computed often using the one-body part of the Coulomb potential. The two-body Coulomb interaction because of its long-range is dominated by the monopole part in the multipole expansion. When considering off-diagonal Coulomb matrix elements, the most important part is the one-body matrix element involving the monopole.
In fact, the isovector monopole matrix element between the ground state and the isovector monopole can be approximately written as [8]:

$$\langle 0|V_c^{(1)}|M\rangle = \frac{1}{7}Z \text{ MeV} \tag{28}$$

which for a nucleus like $^{40}Ca$ is 3 MeV.
The other parts of the two-body Coulomb force cannot contribute much to Coulomb mixing, unless there is an accidental degeneracy between levels that mix. This of course is not the case for the ground state. Coulomb mixing (including isospin mixing) is determined by the distribution of the isovector monopole strength.

In ref. [8] studies of Coulomb mixing were presented in which the notion of the giant isovector monopole was extensively used. The reader is referred to this reference, where various models of isospin mixing via the isovector monopole are described. Here we use the results in [8] in order to calculate expressions for $\delta_c$.

As we deal with $T_z = \pm 1$ nuclei we will take $Z \approx \dfrac{A}{2}$ and express the dependence on $A$ only. Below are presented the results for $\delta_c$ using the four models, for the $T_z = \pm 1$ nuclei.

The first one we employ is the hydrodynamical model of Bohr and Mottelson [9] in which the IVMS is the result of radial oscillations of the proton fluid against the neutron fluid. In this model the energy of the IVMS is very high, $\xi$ being more than 4.

1. The hydrodynamical model:

$$\delta_c = 6.0 \times 10^{-7} A^2 \qquad (29)$$

The next model we apply here is based on the Non-Energy Weighted Sum Rule ((NEWSR).for the isovector monopole strength. For the approximate derivation of this sum rule see ref. [8]. The result for $\delta_c$ is given below.

2. NEWSR

$$\delta_c = 0.67 \times 10^{-7} A^{7/3} \qquad (30)$$

We will also use here the results for the isospin impurity obtained in [8] using the energy weighted sum rule (EWSR) for the isovector monopole resonance. Using the expression for $\varepsilon_1^2$ one finds

3. EWSR

$$\delta_c = 5.7 \times 10^{-7} A^2 \qquad (31)$$

In all three above models the IVMS is treated as a doorway state and one allows for a spreading width of this resonance. (Some small fraction of this strength can reach relatively low energies).

Finally we will use the results of some simple microscopic RPA calculations of the IVMS In which schematic p-h interactions were used. In N>Z nuclei the separation of the two isospin components with $T$ and $T+1$ for the IVMS was taken into account. To obtain good isospin states it is necessary to include certain 2p-2h configurations in the wave function of the IVMS [8]. This was explicitly done and the $T+1$ components of the IVMS were determined. Using these components the isospin admixtures were evaluated for a series of nuclei. A phenomenological formula for the isospin impurity was obtained by fitting these results for several nuclei with different masses [8].

4. Microscopic :

$$\delta_c = 18.0 \times 10^{-7} A^{5/3} \qquad (32)$$

Numerical results for several masses $A$ and the four models are presented in Table 1.

| $\delta_c$ \ A | 10 | 40 | 80 |
|---|---|---|---|
| Hydrodynamical | 0.006% | 0.1% | 0.4% |
| NEWSR | 0.001% | 0.04% | 0.20% |
| EWSR | 0.005% | 0.09% | 0.4% |
| Microscopic | 0.009% | 0.06% | 0.25% |

Table1. Values of $\delta_c$ in % for several mass numbers A for the four models discussed in the text.

## IV DISCUSSION

The spread of values for $\delta_c$ in Table 1 is within a factor of 2 for the various models considered. Comparing these to the results for $\delta_c$ in [1, 2] one sees that our calculation predicts considerably lower values, by factors 2-4. When comparing the calculated $\delta_c$ to the ones obtained in the shell-model [3] one observes that the numbers for $\delta_{IM}$ (Isospin Mixing) from [3] are roughly in agreement with the $\delta_c$ in the present work. However, in [3] the $\delta_{RO}$ correction (termed the "radial overlap") is large and as stated in [1-3] should be added to $\delta_{IM}$ in order to get $\delta_c$. As already emphasized the $\delta_c$ calculated here includes both contributions. Therefore
our results for $\delta_c$ are smaller than those in [3].
  Why is there this difference between the results of our approach and the ones discussed above? It is difficult to pinpoint exactly the reasons; one possible reason is that in the other works collective effects are not included. In the present work on the contrary, the mixing with IVMS takes into account effects of collectivity. The IVMS is a collective excitation and because of the repulsive nature of the particle-hole interaction in the isovector mode it is shifted to higher energies and its strength is reduced. This leads to reduced Coulomb mixing both, in the proton wave function and in the isopin impurity of the isospin quantum number.
    The question of using the analog spin $W$ versus the use of isospin $T$ in some calculations of $\delta_c$ [1, 2] was raised recently in ref. [4]. In the analog spin formalism the $W_-$ operator, for example, changes a neutron in a neutron orbit into a proton occupying the corresponding proton orbit which is distorted by the Coulomb potential. The operator

$T_-$, on the other hand, changes the neutron charge but does not change the orbit. (Analogously for the $W_+$ and $T_+$ operators). As mentioned the operator to be used in eq. (2) is the $T_+$. In our approach we employ consistently the isospin formalism. It is worth noticing the Coulomb admixtures of the IVMS introduce distortions into the proton single particle wave functions and, as has been demonstrated, this mixing in the isobaric analog state is equivalent to the formation of a $W$-analog state. In our approach this is a result of the calculation and not the starting point. The so called "radial overlap correction" is included in a consistent manner, avoiding double counting, without the need to use the $W$-spin.

It is clearly exhibited in our approach that the correction $\delta_c$ depends explicitly on two quantities; the isospin impurity and the strength of the symmetry potential. Smaller symmetry potential leads to a smaller $\delta_c$ correction.

It is important to asses the uncertainties in our treatment of $\delta_c$. As already mentioned the symmetry potential strength is not well determined. This parameter determines the splitting between the various isospin components. There is another factor that influences this splitting, namely the different degree of collectivity of these isospin components. In large neutron excess nuclei this might alter considerably the spacing [8], but in $T_z = \pm 1$ nuclei the collectivity of the various components of the IVMS is similar and the effect on the spacing is small. The IVMS has a spreading width and this could bring some fraction of strength to lower energies and influence the result. Also the centroid energy represented by the factor $\xi$ contains some degree of uncertainty. The use of a simplified charge distribution (homogenous sphere) and the neglect of short-range non-Coulomb charge-dependent interactions might affect the results somewhat. There are possibly a number of other small uncertainties. If we rely on an intuitive estimate that the maximal uncertainty is 50%, still our results for $\delta_c$ are considerably lower than the ones found in previous studies. However, the effect of these reduced values on $V_{ud}$ will not be that strong because the radiative corrections (which we do not treat here) dominate.

## ACKNOWLEDGEMENTS

We wish to thank Byron Jennings for his hospitality at TRIUMF where this work was performed.